\documentclass[12pt]{article}

\usepackage[latin1]{inputenc}
\usepackage{amssymb}
\usepackage{amsfonts} 
\usepackage{graphicx} 
\usepackage{amsthm} 
\usepackage{amsmath} 
\usepackage{epsfig}
\usepackage{hyperref}

\def\nn{\nonumber}

\def \bc {\begin{center}}
\def \ec {\end{center}}
\def \bi {\begin{itemize}}
\def \ei {\end{itemize}}

\def \ba {\begin{array}}
\def \ea {\end{array}}

\def \bea {\begin{eqnarray}}
\def \eea {\end{eqnarray}}

\def \be {\begin{equation}}
\def \ee {\end{equation}}

 \def\mbN {{\mathbb N}}

\def \lb {\left[}
\def \rb {\right]}

\def \la {\langle}
\def \ra {\rangle}

\def \cn {N}
\def \um {\frac{1}{2}}
\def \bz {\bar{z}}

\def\cD {{\cal D}}

\def\tr {{\rm tr}}

\newcommand{\ld}{\rangle}

\theoremstyle{remark}

\begin{document}

\begin{center}
{\LARGE {\bf Coherent States of Accelerated Relativistic Quantum Particles, Vacuum Radiation and the Spontaneous Breakdown of the Conformal SU(2,2) Symmetry}}
\end{center}
\bigskip
\bigskip

\centerline{{\sc M. Calixto}$^{1,2}$\footnote{Corresponding author:
calixto@ugr.es}, {\sc E. Pérez-Romero}$^{1}$ and {\sc V.
Aldaya}$^{2}$ }

\bigskip

\bc {\it $^1$ Departamento de Matemática Aplicada,
Universidad de Granada,  Facultad de Ciencias, Campus de Fuentenueva, 18071 Granada, Spain}
\\
{\it $^2$ Instituto de Astrof\'\i sica de Andaluc\'\i a (IAA-CSIC),
Apartado Postal 3004, 18080 Granada, Spain}

 \ec

\bigskip
\begin{center}
{\bf Abstract}
\end{center}
\small
\begin{list}{}{\setlength{\leftmargin}{3pc}\setlength{\rightmargin}{3pc}}
\item 
We give a quantum mechanical description of accelerated relativistic particles in the framework of Coherent States (CS) of the 
(3+1)-dimensional conformal group $SU(2,2)$, with the role of accelerations played by special conformal transformations and with the 
role of (proper) time translations played by dilations. 
The accelerated ground state $\tilde\varphi_0$ of first quantization is a CS of the conformal group. We compute the distribution 
function giving the occupation number of each energy level in $\tilde\varphi_0$ and, with it, the partition function ${\cal Z}$, mean energy ${\cal E}$ 
and entropy ${\cal S}$, which resemble that of an ``Einstein Solid''. An effective temperature 
${\cal T}$ can be assigned to this ``accelerated ensemble'' through the thermodynamic expression $d{\cal E}/d{\cal S}$, which leads to a (non linear) relation 
between acceleration and temperature different from Unruh's (linear) formula.  Then we construct the corresponding 
conformal-$SU(2,2)$-invariant second quantized theory and its spontaneous
breakdown when selecting Poincaré-invariant degenerated $\theta$-vacua (namely,
coherent states of conformal zero modes). Special conformal
transformations (accelerations) destabilize the Poincaré vacuum and make
it to radiate.
\end{list}
\normalsize 





\newpage

\section{Introduction}

The quantum analysis of accelerated frames of reference has been studied mainly in connection with
Quantum Field Theory (QFT) in curved space-time.  For example, the case of
the quantization of a Klein-Gordon field in Rindler coordinates (see e.g. \cite{Davies,Birrell} and Appendix \ref{mutilation} for a short review)  entails a
global mutilation of flat space-time, with the appearance of event horizons, and
leads to a quantization inequivalent to the standard Minkowski quantization. Physically one says that, whereas the
Poincaré-invariant (Minkowskian) vacuum $|0\rangle$ in QFT looks the same to any
inertial observer (i.e., it is stable under Poincaré transformations),
it converts into a thermal bath of radiation with temperature
\be T=\frac{\hbar a}{2\pi vk_B}\label{tempunruh}\ee
in passing to a uniformly accelerated frame ($a$ denotes the acceleration,
$v$ the speed of light\footnote{In this paper, the letter $c$ is reserved
for special conformal transformations (relativistic uniform
accelerations)} and $k_B$ the Boltzmann constant). This is called the Fulling-Davies-Unruh effect
\cite{Davies,Fulling,Unruh}, which shares some features with the
(black-hole) Hawking \cite{Hawking} effect. Its explanation relies heavily upon Bogoliubov transformations, which find 
a natural explanation in the framework of Coherent States (CS) and squeezed states \cite{Bishop}.

In this article we also approach the quantum analysis of accelerated frames from a CS perspective but the scheme is rather different, although 
it shares some features with the standard approach commented before.  The situation will be similar in some respects to quantum many-body 
condensed mater systems describing, for example, superfluidity and superconductivity, where the ground state mimics the quantum 
vacuum in many respects and quasi-particles (particle-like excitations above the ground state) play the role of matter. We shall enlarge 
the Poincaré symmetry ${\cal P}$ to account for uniform accelerations and then we shall spontaneously break it down back to Poincaré by selecting 
appropriate ``non-empty vacua''\footnote{Actually, quantum vacua are not really empty to every observer, as the quantum vacuum is filled with zero-point fluctuations of
quantum fields.} stable under ${\cal P}$. Then the action of broken symmetry transformations (accelerations) will 
destabilize/excitate the vacuum and make it to \emph{radiate}. The candidate for an enlargement of  ${\cal P}$ will be 
the conformal group in (3+1)-dimensions $SO(4,2)$ incorporating dilations and  special
conformal transformations (STC) 
\be x^\mu \rightarrow {x'}^\mu=\frac{x^\mu+c^\mu x^2}{1+2c x+c^2
x^2}\label{confact-1}, \ee
which can be interpreted as transitions to systems of relativistic,
uniformly accelerated observers with acceleration $a=2c$ (see e.g. Ref.
\cite{Hill,Cervero1975,conforme-ac} and later on Eq. (\ref{sctrans1})). From the conformal symmetry point of view, 
Poincaré-invariant vacua are regarded as a coherent states of \emph{conformal zero modes}, which are
undetectable (``dark'') by inertial observers but unstable under special
conformal transformations. 

A previous preliminary attempt to analyze quantum accelerated frames from a conformal group perspective was made 
in the reference \cite{conforme} (see also \cite{CQG}), where a quite involved ``second quantization
formalism on a group $G$'' was developed and applied  to the (finite part of the)
conformal group in (1+1) dimensions, $SO(2,2)\simeq SO(2,1)\times
SO(2,1)$, which consists of two copies of the pseudo-orthogonal group
$SO(2,1)$ (left- and right-moving modes, respectively). Here we shall use
more conventional methods of quantization and we shall work in realistic
(3+1) dimensions, using the (more involved) conformal group $SO(4,2)\simeq
SU(2,2)/\mathbb Z_4$. New consequences of this group-theoretical approach are obtained here,
regarding a similitude between the accelerated ground state and the ``Einstein Solid'', the computation of entropies and 
a deviation from the Unruh's formula (\ref{tempunruh}).

We would like to mention that (near-horizon \emph{two}-dimensional) conformal symmetry has also played a fundamental role in the
microscopic description of the Hawking effect. In fact, there is strong evidence that conformal field theories provide a universal 
(independent of the details of the particular quantum gravity model) description of low-energy black hole entropy, which
is only fixed by symmetry arguments (see e.g. \cite{Carlipcqg,Agullo}). Here, the Virasoro algebra turns out to be the
relevant subalgebra of surface deformations of the horizon of an arbitrary
black hole and constitutes the general gauge (diffeomorphism) principle that governs the density of states.
However, in 3+1 dimensions,  conformal
invariance is necessarily global (finite-(15)-dimensional). In this paper we shall study zero-order effects
that gravity has on quantum theory (uniform accelerations). To account for higher-order effects
(like non-constant accelerations) in a group-theoretical framework, we should firstly promote the 3+1
conformal symmetry SO(4,2) to a higher-(infinite)-dimensional symmetry. This is not a trivial task, although some steps have
been done by the authors in this direction (see e.g. \cite{infdimal,promoting,CQG,gp,jarama}).

The organization of the paper is as follows.  In Section \ref{gtb} we discuss the group theoretical backdrop
(conformal transformations, infinitesimal generators and commutation relations) and justify the
interpretation of special conformal transformations as transitions to relativistic
uniform accelerated frames of reference. In
Section  \ref{mcqp} we construct the Hilbert space and an orthonormal basis for our conformal particle
in 3+1 dimensions, based on an holomorphic square-integrable irreducible representation of the conformal group on
the eight-dimensional phase space $\mathbb D_4=SO(4,2)/SO(4)\times SO(2)$ inside the complex Minkowski space $\mathbb C^4$. In Section
\ref{breakdown} we remind the general definition of CS of a group $G$, highlight the Poincaré invariance of the ground state, construct 
the accelerated ground state as a CS of the conformal group and calculate the distribution function, mean energy, partition function and
entropy of this accelerated ground state, seen as a statistical ensemble. This leads us to interpret the accelerated ground state as an Einstein Solid, to obtain
a deviation from the Unruh's formula (\ref{tempunruh}) and to discuss on the existence  of a
maximal acceleration. In Section \ref{thetasec} we deal with the second-quantized (many-body) theory, where Poincaré-invariant (degenerated)
pseudo-vacua  are coherent states of conformal zero modes. Selecting one of this Poincaré-invariant pseudo-vacua spontaneously breaks
the conformal invariance and leads to vacuum radiation. Section \ref{conclu} is left for conclusions and outlook. In addition, two appendices 
are also included. Appendix \ref{mutilation} reminds Rindler coordinates and Bogoliubov transformations in the standard derivation of Unruh effect and 
Appendix \ref{conf-sigma}  reports on the underlying  gauge-invariant Lagrangian formalism behind our quantum model of conformal particles.

\section{The Conformal Group and its Generators}\label{gtb}

The conformal group in 3+1 dimensions, $SO(4,2)$,  is composed by
Poincaré ${\cal P}=SO(3,1)\circledS \mathbb R^4$ [semidirect product of spacetime translations $b^\mu\in\mathbb R^4$ times Lorentz
$\Lambda^{\mu}_\nu\in SO(3,1)$] transformations augmented by dilations
($e^\tau\in\mathbb R_+$) and relativistic uniform accelerations (special
conformal transformations, $c^\mu\in\mathbb R^4$) which, in Minkowski
spacetime, have the following realization:
\be\ba{rcl} x'^\mu &=& x^\mu+b^\mu,\;\;x'^\mu=\Lambda^{\mu}_\nu(\omega) x^\nu,\\
x'^\mu&=&e^\tau x^\mu,\;\;x'^\mu=\frac{x^\mu+c^\mu x^2}{1+2c x+c^2
x^2},\ea\label{confact} \ee
respectively. The infinitesimal generators (vector fields) of the
transformations (\ref{confact}) are easily deduced:
\be \ba{rcl}P_\mu &=& \frac{\partial}{\partial x^\mu}, \;\;
M_{\mu\nu}=x_\mu
\frac{\partial}{\partial x^\nu}-x_\nu \frac{\partial}{\partial x^\mu},\\
D&=&x^\mu\frac{\partial}{\partial x^\mu},\;\; K_\mu=-2x_\mu x^\nu
\frac{\partial}{\partial x^\nu}+x^2\frac{\partial}{\partial x^\mu}
\label{confvf}\ea\ee
and they close into the conformal Lie algebra
 \be\ba{rcl} \lb M_{\mu\nu},M_{\rho\sigma}\rb &=&\eta_{\nu\rho}M_{\mu\sigma}+\eta_{\mu\sigma}M_{\nu\rho}
-\eta_{\mu\rho}M_{\nu\sigma}-\eta_{\nu\sigma}M_{\mu\rho},\\
\left[P_\mu,M_{\rho\sigma}\right] &=& \eta_{\mu\rho} P_\sigma -
\eta_{\mu\sigma}
P_\rho,\;\; \lb P_\mu,P_\nu\rb=0,\\
\lb K_\mu,M_{\rho\sigma}\rb &=& \eta_{\mu\rho}K_\sigma-\eta_{\mu\sigma}K_\rho,\;\; \lb K_\mu,K_\nu\rb=0, \\
\lb D,P_\mu\rb &=&-P_\mu, \;\;\lb D,K_\mu\rb =K_\mu,\;\; \lb D,M_{\mu\nu}\rb=0,\\
\lb K_\mu,P_\nu\rb &=& 2(\eta_{\mu\nu}
D+M_{\mu\nu}).\ea\label{conformalgebra}\ee
The conformal quadratic Casimir operator
\be C_2=D^2-\um M_{\mu\nu}M^{\mu\nu}+\um(P_\mu K^\mu+K_\mu P^\mu),\label{Casimir}\ee
generalizes the Poincaré Casimir $P^2=P_\mu P^\mu$  which, for scalar fields $\phi$, leads to the Klein-Gordon equation
$P^2\phi=m^2_0\phi$, with $m^2_0$ the squared rest mass. The fact that $[D,P^2]=-2P^2$ implies that conformal fields 
must be either massless or to have a continuous mass spectrum (see e.g. the classical Refs. \cite{Barutmass} and \cite{CMP55}). 
Actually, just like the Poincaré invariant mass $m_0$  comprises a continuum of ``Galilean'' masses $m$,  a conformally invariant mass 
$m_{00}$ can be defined by the Casimir (\ref{Casimir}), which comprises a continuum of Poincaré masses $m_0$. The eigenvalue 
equation  $C_2\phi=m_{00}^2\phi$ can be seen as a \emph{generalized} Klein-Gordon
equation, where $D$ replaces $P_0$ as the
(proper) time evolution  generator and $m_{00}$  replaces $m_0$ 
(see Appendix \ref{conf-sigma} for more information and \cite{Barutmass} for the formulation of other
conformally-invariant massive field equations of motion in
generalized Minkowski space).

In this article we shall deal with discrete series representations 
of the conformal group having continuous mass spectrum and the corresponding 
wavefunctions having support on the whole four-dimensional Minkowski space-time, 
with the dilation parameter $\tau$ playing the role of a proper time.  We shall report on this model of conformal quantum 
particles later on Sec. \ref{mcqp}.  The reader can also consult 
our recent reference \cite{spinning-part}  for a gauge-invariant Lagrangian approach (of nonlinear
sigma-model type),  using a generalized Dirac method for the quantization of constrained systems. We give here a flavor of 
this approach in the  Appendix \ref{conf-sigma}.

\subsection{Special conformal transformations as transitions to uniform relativistic
accelerated frames}

The interpretation of special conformal transformations (\ref{confact-1})
as transitions from inertial reference frames to systems of relativistic,
uniformly accelerated observers was identified many years ago by
\cite{Hill,Cervero1975,conforme-ac}. More precisely, denoting by
$u^\mu=\frac{dx^\mu}{d\tau}$ and $a^\mu=\frac{du^\mu}{d\tau}$ the
four-velocity and four-acceleration of a point particle, respectively, the
relativistic motion with constant acceleration is characterized by the
usual condition \cite{Misner-Thorne-Wheeler}:
\be a_\mu a^\mu=-{\rm g}^2,\label{constanta}\ee
where ${\rm g}$ is the magnitude of the acceleration in the instantaneous
rest system. From $u_\mu u^\mu=1$ (in $v=1$ unities) and
(\ref{constanta}), we can derive the differential equation to be satisfied
for all systems with constant relative acceleration\footnote{As a
curiosity, this formula turns out to be equivalent to the vanishing of
the von Laue four-vector $F^\mu=\frac{2}{3}e^2(\frac{da^\mu}{d\tau}+a_\nu
a^\nu u^\mu)$ of an accelerated point charge; that is, a compensation
between the Schott term $\frac{2}{3}e^2\frac{da^\mu}{d\tau}$ and the
Abraham-Lorentz-Dirac radiation reaction force $\frac{2}{3}e^2a_\nu a^\nu
u^\mu$ (minus the rate at which energy and momentum is carried away from
the charge by radiation)}:
\be \frac{da^\mu}{d\tau}={\rm g}^2u^\mu. \label{confeq}\ee
Hill \cite{Hill} (see also \cite{Cervero1975} and \cite{conforme-ac})
proved that the kinematical invariance group of (\ref{confeq}) is
precisely the conformal group $SO(4,2)$. Here we shall provide a simple
explanation of this fact. For simplicity, let us take an acceleration
along the ``z'' axis: $c^\mu=(0,0,0,c)$, and the temporal path
$x^\mu=(t,0,0,0)$. Then the transformation (\ref{confact-1}) reads:
\be t'=\frac{t}{1-c^2t^2}\;,\;\;\;\;\;\; z'=\frac{c 
t^2}{1-c^2t^2}.\label{sctrans1}\ee
Writing $z'$ in terms of $t'$ gives the usual formula for the relativistic
uniform accelerated (hyperbolic) motion: \be z'
=\frac{1}{a}(\sqrt{1+{a^2t'^2}}-1) \ee with $a=2c$.

Let us say that at least two alternative meanings of
special conformal transformations (STC) have also been proposed \cite{Weyl,Kastrup1}. One is related to the
Weyl's idea of different lengths in different points of space time
\cite{Weyl}: ``the rule for measuring distances changes at different
positions''. Other is Kastrup's interpretation of SCT as geometrical gauge
transformations of the Minkowski space \cite{Kastrup1}.

\section{A Model of Conformal Quantum Particles}\label{mcqp}

In this Section we report on a model for quantum particles with conformal symmetry. The reader can find more details in the Reference \cite{spinning-part}, where
we formulate a gauge invariant nonlinear sigma-model on the conformal group and quantize it according to a generalized Dirac method for constrained systems.

\subsection{The compactified Minkowski space and the isomorphism $SO(4,2)=SU(2,2)/\mathbb Z_4$}

In \cite{spinning-part} (see also the Appendix \ref{conf-sigma}) it is shown how the Minkowski space arises as the support of constrained wave functions on the 
conformal group. Actually, the compactified Minkowski space $\mathbb
M_4=\mathbb S^3\times_{\mathbb Z_2} \mathbb S^1$ naturally lives inside the conformal group $SO(4,2)$ as the coset 
 $\mathbb M_4=SO(4,2)/{\cal W}$, where ${\cal W}$
denotes the Weyl subgroup generated by $K_\mu, M_{\mu\nu}$ and $D$ (i.e.,
a Poincaré subgroup ${\cal P}$ augmented by
the dilations $\mathbb R^+$). The Weyl group ${\cal W}$ is the stability
subgroup (the little group in physical usage) of $x^\mu=0$. The
conformal group acts transitively on $\mathbb M_4$ and free from
singularities.

Instead of $SO(4,2)$, we shall work by convenience with  its four covering group:
\be SU(2,2)=\left\{g=\left(\ba{cc} A& B
\\ C &D\ea\right)\in {\rm Mat}_{4\times 4}(\mathbb C):  g^\dag \Gamma g=\Gamma,
\det(g)=1\right\}, \label{su22def}\ee
where $\Gamma$ denotes a hermitian form of signature $(++--)$.

The conformal Lie algebra (\ref{conformalgebra}) can also be realized  in
terms of gamma matrices in, for instance, the Weyl basis
\be \gamma^\mu=\left(\ba{cc} 0& \sigma^\mu \\ \check{\sigma}^\mu
&0\ea\right),\;\; \gamma^5=i\gamma^0\gamma^1\gamma^2\gamma^3=\left(\ba{cc}
-\sigma^0& 0\\ 0& \sigma^0\ea\right),\ee
where $\check{\sigma}^\mu\equiv \sigma_\mu$ (we are using the convention
$\eta={\rm diag}(1,-1,-1,-1)$) and $\sigma^\mu$ are the standard Pauli
matrices
\be \sigma^0=I=\left(\ba{cc} 1& 0
\\ 0 &1\ea\right),\;\sigma^1=\left(\ba{cc} 0& 1
\\ 1 &0\ea\right),\;\sigma^2=\left(\ba{cc} 0& -i
\\ i &0\ea\right),\;\sigma^3=\left(\ba{cc} 1& 0
\\ 0 &-1\ea\right).\ee
Indeed, the choice
\bea D&=&\frac{\gamma^5}{2},\;M^{\mu\nu}=\frac{\lb
\gamma^\mu,\gamma^\nu\rb}{4}=\frac{1}{4}\left(\ba{cc} \sigma^\mu\check{\sigma}^\nu-\sigma^\nu\check{\sigma}^\mu & 0\\
0&\check{\sigma}^\mu\sigma^\nu-\check{\sigma}^\nu\sigma^\mu\ea\right),\nn\\
P^\mu&=&\gamma^\mu\frac{1+\gamma^5}{2}=\left(\ba{cc} 0& \sigma^\mu
\\ 0
&0\ea\right),\;K^\mu=\gamma^\mu\frac{1-\gamma^5}{2}=\left(\ba{cc} 0& 0
\\ \check{\sigma}^\mu &0\ea\right)
\label{confalgamma}\eea
fulfills the commutation relations (\ref{conformalgebra}). These are the
Lie algebra generators of the fundamental representation of $SU(2,2)$.

The group $SU(2,2)$  acts transitively on the compactified Minkowski space
$\mathbb M_4$, which can be identified with the set of hermitian $2\times 2$ matrices $X=x_\mu\sigma^\mu$, as follows:
\be X\to X'=(AX+B)(CX+D)^{-1}.\label{tubeaction}\ee
With this identification, the transformations
(\ref{confact}) can be recovered from (\ref{tubeaction}) as follows:
\begin{itemize}
 \item[i)] Standard Lorentz transformations, $x'^\mu=\Lambda^{\mu}_\nu(\omega) x^\nu$, correspond to $B=C=0$ and
$A=D^{-1\dag}\in SL(2,\mathbb C)$, where we are making use of the
homomorphism (spinor map) between $SO^+(3,1)$ and $SL(2,\mathbb C)$ and
writing $X'=AXA^\dag, A\in SL(2,\mathbb C)$ instead of
$x'^\mu=\Lambda^{\mu}_\nu x^\nu$. \item[ii)] Dilations correspond to
$B=C=0$ and $A=D^{-1}=e^{\tau/2}I$ \item[iii)] Spacetime translations
are $A=D=I$, $C=0$ and $B=b_\mu\sigma^\mu$.
\item[iv)] Special conformal transformations correspond to $A=D=I$
and $C=c_\mu\sigma^\mu, B=0$ by noting that $\det(CX+I)=1+2cx+c^2x^2$: \be
X'=X(CX+I)^{-1}\leftrightarrow x'^\mu=\frac{x^\mu+c^\mu x^2}{1+2cx+c^2
x^2}\nn\ee
\end{itemize}

\subsection{Unirreps of the conformal group: discrete series}

We shall consider the complex extension of the compactified Minkowski space $\mathbb M_4=U(2)$ to
the 8-dimensional conformal (phase) space:
\be \mathbb
D_4=U(2,2)/U(2)^2=\{Z\in {\rm Mat}_{2\times 2}(\mathbb C):
I-ZZ^\dag>0\},\ee
of which $\mathbb M_4=\{Z\in {\rm Mat}_{2\times 2}(\mathbb C):
I-ZZ^\dag=0\}$ is the Shilov boundary. It can be proved (see e.g.  \cite{spinning-part} and \cite{MacMahon}) 
that the following action
\be [U_\lambda(g)\phi](Z)=|CZ+D|^{-\lambda}\phi(Z'),\;
Z'=(AZ+B)(CZ+D)^{-1}\label{repre}\ee
constitutes a unitary irreducible representation of $SU(2,2)$ on the Hilbert space
${\cal H}_\lambda(\mathbb D_4)$ of square-integrable holomorphic
functions $\phi$ with invariant integration measure
\be
d\mu_\lambda(Z,Z^\dag)=\pi^{-4}(\lambda-1)(\lambda-2)^2(\lambda-3)
\det(I-ZZ^\dag)^{\lambda-4} |dZ|,\nn\ee
where the label $\lambda\in\mathbb Z, \lambda\geq 4$ is the conformal,
scale or mass dimension ($|dZ|$ denotes the Lebesgue measure in $\mathbb
C^4$). The factor $\pi^{-4}(\lambda-1)(\lambda-2)^2(\lambda-3)$ in $d\mu_\lambda(Z,Z^\dag)$ is chosen so that the 
constant function $\phi(Z)=1$ has unit norm. 
Besides the conformal dimension $\lambda$, the discrete series representations of $SU(2,2)$ have
two extra spin labels  $s_1,s_2\in\mathbb N/2$ associated with the (stability) subgroup $SU(2)\times SU(2)$. Here
we shall restrict ourselves to  scalar fields ($s_1=s_2=0$) for the sake of
simplicity (see e.g. \cite{spinning-part} for the spinning unirreps of $SU(2,2)$). The reduction of this 
representation  into unitary irreducible representations of the Poincaré subgroup indicates that 
we are dealing with fields with a continuous mass spectrum extending from zero to infinity \cite{Ruhl}.

\subsection{The Hilbert space of our conformal particle}

It has been proved in \cite{MacMahon} that the infinite set of homogeneous
polynomials
\be
\varphi_{q_1,q_2}^{j,m}(Z)=\sqrt{\frac{2j+1}{\lambda-1}\binom{m+\lambda-2}{\lambda-2}\binom{m+2j+\lambda
-1}{\lambda-2}}\det(Z)^{m}\cD^{j}_{q_1,q_2}(Z),\label{polibase}\ee
with \be
\cD^{j}_{q_1,q_2}(Z)=\sqrt{\frac{(j+q_1)!(j-q_1)!}{(j+q_2)!(j-q_2)!}}
\sum_{p=\max(0,q_1+q_2)}^{\min(j+q_1,j+q_2)}\tbinom{j+q_2}{p}\tbinom{j-q_2}{p-q_1-q_2}
z_{11}^p
z_{12}^{j+q_1-p}z_{21}^{j+q_2-p}z_{22}^{p-q_1-q_2}\label{Wignerf}\ee
the standard Wigner's $\cD$-matrices ($j\in\mbN/2$), verifies the
following closure relation (the reproducing Bergman kernel or $\lambda$-extended MacMahon-Schwinger's master formula):
\be\sum_{j\in\mbN/2}\sum^{\infty}_{m=0}\sum^{j}_{q_1,q_2=-j}
\overline{\varphi_{q_1,q_2}^{j,m}({Z})}\varphi_{q_1,q_2}^{j,m}(Z')=\frac{1}{\det(I-Z^\dag
Z')^\lambda}\label{Bergman}\ee
and constitutes an orthonormal basis of ${\cal H}_\lambda(\mathbb D_4)$
(the sum on $j$ accounts for all non-negative half-integer numbers). The identity (\ref{Bergman}) will be
usefull for us in the sequel.

\subsection{Hamiltonian and energy spectrum}\label{hamilsec}

In \cite{spinning-part} we have argued that the dilation operator $D$
plays the role of the Hamiltonian of our conformal quantum theory. Actually, the
replacement of time translations by dilations as kinematical equations of
motion has already been considered in the literature (see e.g. \cite{conformecontract} and in
\cite{dilatatiempo}), when quantizing field theories on space-like
Lorentz-invariant hypersurfaces $x^2=x^\mu x_\mu=\tau^2=$constant. In
other words, if one wishes to proceed from one surface at $x^2=\tau_1^2$
to another at $x^2=\tau_2^2$, this is done by scale transformations; that
is, $D=\frac{\partial}{\partial \tau}$ is the evolution operator in a proper time $\tau$. 
We must say that other possibilities exist for choosing a conformal Hamiltonian, namely the combination $\tilde P_0=(P_0+K_0)/2$,
which has been used in \cite{CMP55}.

From the general expression (\ref{repre}), we can compute the finite
left-action of dilations  ($B=0=C$ and
$A=e^{\tau/2}\sigma^0=D^{-1} \Rightarrow g=e^{\tau/2}{\rm diag}(1,1,-1,-1)$) on wave functions,
\be [U_\lambda(g)\phi](Z)=e^{\lambda\tau}\phi(e^\tau Z).\ee
The infinitesimal generator of this transformation is the Hamiltonian
operator:
\be H=\lambda+\sum_{i,j=1}^2Z_{ij}\frac{\partial}{\partial
Z_{ij}}=\lambda+z_\mu\frac{\partial}{\partial{z_\mu}},\label{Hamiltonian}\ee
where we have set $Z=z_\mu\sigma^\mu$ in the last equality. This Hamiltonian has the form
of that of a four-dimensional (relativistic) harmonic oscillator in the Bargmann representation.
The set of functions (\ref{polibase}) constitutes a basis of  Hamiltonian
eigenfunctions (homogeneous polynomials) with energy eigenvalues
${E}_{n}^\lambda$ (the homogeneity degree) given by:
\be H\varphi_{q_1,q_2}^{j,m}={E}_{n}^\lambda\varphi_{q_1,q_2}^{j,m},\;\;
{E}_{n}^\lambda=\lambda+n,\;\; n=2j+2m.\label{energyspectrum}\ee
Actually, each energy level ${E}_{n}^\lambda$ is $(n+1)(n+2)(n+3)/6$ times
degenerated (just like a four-dimensional harmonic oscillator). This degeneracy coincides with the number of linearly
independent polynomials $\prod_{i,j=1}^2 Z_{ij}^{n_{ij}}$ of fixed degree
of homogeneity $n=\sum_{i,j=1}^2n_{ij}$. This also proves that the set of
polynomials (\ref{polibase}) is a basis for analytic functions $\phi\in
{\cal H}_\lambda(\mathbb D_4)$.  The spectrum is equi-spaced and bounded
from below, with ground state $\varphi_{0,0}^{0,0}=1$ and zero-point
energy $E_{0}^\lambda=\lambda$ (the conformal, scale or mass dimension).

\section{Coherent States of Accelerated Relativistic Particles, Distribution Functions and Mean Values} \label{breakdown}

Before introducing coherent states of $G=SU(2,2)$, let us briefly remind some general definitions and constructions for a
general group $G$. More information on coherent states can be found in standard text books like \cite{Perelomov,Klauder,Gazeau}.

\subsection{A brief on coherent states}

Essential ingredients to define and construct Coherent States (CS) on a given symmetry (Lie) group $G$ are the following.
Firstly, we need a \emph{unitary} representation $U$ of $G$ on
a Hilbert space $({\cal H},\langle \cdot|\cdot\rangle)$. Consider also the
space $L^2(G,dg)$ of square-integrable complex functions $\Psi$ on $G$, where
$dg=d(g'g),\,\forall g'\in G$, stands for the left-invariant
Haar measure on $G$. A non-zero
vector $\varphi^0\in {\cal H}$ is called \emph{admissible}  (or a
\emph{fiducial} vector) if $\Phi(g)\equiv \langle
U(g)\varphi^0|\varphi^0\rangle\in L^2(G,dg)$. A unitary representation $U$ for which admissible vector exists is
called \textit{square integrable}. Assuming that the representation $U$ is \emph{irreducible}, and that
there exists a function $\varphi_0$ admissible, then a system of CS of
${\cal H}$  associated to (or indexed by) $G$ is defined
as the set of functions in the orbit of $\varphi^0$ under $G$:
\be
\varphi_g^0\equiv U(g)\varphi^0, \;\; g\in G,
\ee
and they form an overcomplete set in ${\cal H}$. We can also restrict ourselves to a suitable
homogeneous space $Q=G/H$, for some closed subgroup $H$, by taking a convenient Borel section $\sigma:Q\to G$. In this case, the
set of CS $\{\varphi_{\sigma(q)}^0, q\in Q\}$ is indexed by points in $Q$.

The best known example of CS are ``canonical'' CS associated to the Heisenberg-Weyl group, with Lie algebra commutation
relations $[\hat a,\hat a^\dag]=1$  in terms of annihilation (lowering $\hat a$)
and creation (rising $\hat a^\dag$) ladder operators. The Hilbert space ${\cal H}$ is spanned by the (normalized) eigenstates
$|n\rangle$, $n=0,1,2,\dots$, of the (Hermitian) number operator $\hat\cn=\hat a^\dag \hat a$.
These states can be generated from the the Fock vacuum  $|0\rangle$
as:
\be|n\rangle=\frac{1}{\sqrt{n!}}(\hat a^\dag)^n|0\rangle.\label{Fockbasis}\ee
The Fock vacuum  $|0\rangle$ is an admissible vector and a set of CS $\{|z\rangle, \,z\in \mathbb C\}$
are generated by acting with the (unitary)
displacement operator $U(z,\bar{z})\equiv e^{z \hat a^\dagger-\bar{z}
\hat a}=e^{-z\bar{z}/2}e^{z \hat a^\dagger}e^{\bar z \hat a}$ on  $|0\ld$ as follows:
\be |z\ld\equiv
U(z,\bz)|0\ld=e^{-z\bar{z}/2}e^{z\hat a^\dagger}|0\ld=e^{-z\bar{z}/2}\sum_{n=0}^\infty
\frac{z^n}{\sqrt{n!}}|n\rangle.\label{exp-ansion}\ee
They turn out to be eigenstates of the annihilation operator $\hat a$, i.e. $\hat a|z\ld=z|z\ld$. The probability amplitude of finding $n$ quanta 
(namely, photons) in $|z\ld$ is $\varphi_n(z)=\langle n|z\rangle=e^{-|z|^2/2}\frac{z^n}{\sqrt{n!}}$, so that the 
distribution function
\be f_n(|z|)=|\varphi(z)|^2=e^{-|z|^2}\frac{|z|^{2n}}{n!},\ee
is Poissonian, with $|z|^2=\langle z|\hat\cn|z\rangle$ the mean number of ``photons'' in $|z\rangle$. In the next two subsections we shall compute the distribution function
and mean values for CS of accelerated relativistic quantum particles.

\subsection{Conformal CS and the accelerated ground state}

Among the infinite set $\{\varphi_{q_1,q_2}^{j,m}(Z)\}$ of  homogeneous polynomials (\ref{polibase}), we shall
choose the ground state $\varphi^{0,0}_{0,0}(Z)=1$ (of zero degree/energy) as an admissible vector (see \cite{MacMahon} for a
proof of admissibility). The set of CS in the orbit of $\varphi^{0,0}_{0,0}$ under the action (\ref{repre}) are:
\be
\tilde\varphi^{0,0}_{0,0}(Z)=[U_\lambda(g)\varphi_{0,0}^{0,0}](Z)=\det(CZ+D)^{-\lambda}.\label{acvac}\ee
Note that Poincaré transformations (zero acceleration $C=0$ and $\det(D)=1$) leave
the ground state  invariant, that is, $\varphi_{0,0}^{0,0}$ looks the same to
every inertial observer.  We shall call $\tilde\varphi^{0,0}_{0,0}$ the ``accelerated'' ground state. For arbitrary
accelerations, $C=c_\mu\sigma^\mu\not=0$, we can decompose $\tilde\varphi^{0,0}_{0,0}$ using the Bergman kernel expansion
(\ref{Bergman}) as:
\bea \tilde
\varphi_{0,0}^{0,0}(Z)&=&\det(D)^{-\lambda}\det(D^{-1}CZ+I)^{-\lambda}\nn\\ &=&\det(D)^{-\lambda}\sum_{j\in\mbN/2}\sum^{\infty}_{m=0}\sum^{j}_{q_1,q_2=-j}
{\varphi_{q_2,q_1}^{j,m}(-{\cal C})}\varphi_{q_1,q_2}^{j,m}(Z),\label{acvac1}\eea
where ${\cal C}\equiv D^{-1}C$ is a ``rescaled acceleration matrix''. From (\ref{acvac1}), we interpret the coefficient ${\varphi_{q_2,q_1}^{j,m}(-{\cal C})}$ as
the probability amplitude of finding the accelerated ground state in the
excited level $\varphi_{q_1,q_2}^{j,m}$ of energy
$E^\lambda_n=\lambda+2j+2m=\lambda+n$ (up to a global normalizing factor $\det(D)^{-\lambda}$). In the second-quantized
(many particles) theory, the squared modulus $|{\varphi_{q_2,q_1}^{j,m}(-{\cal C})}|^2$ gives us
the occupation number of the corresponding state (see later on Sec. \ref{thetasec}).

\subsection{The accelerated ground state as an statistical ensemble: ``the Einstein solid''}

For canonical ensembles, the (discrete) energy levels $E_n$ of a quantum system in contact with a thermal bath at temperature $T$ are ``populated''
according to the Boltzmann distribution function  $f_n(T)\sim e^{-E_n/k_BT}$. For other external reservoirs or interactions (like, for instance, electric and
magnetic fields acting on a charged particle) one could also compute (in principle) the distribution
function giving the population of each energy level. Actually, if one were able to unitarily implement the external interaction in the original
quantum system, then one could deduce the distribution function for the population of each energy level
from first quantum mechanical principles. This is precisely what we have done with uniform accelerations of Poincaré invariant relativistic
quantum particles, where the unitary transformation (\ref{acvac1}) gives the population of each energy level $E^\lambda_n$ in the accelerated ground state
$\tilde\varphi_{0,0}^{0,0}$.

Let us consider then the coherent state
(\ref{acvac}) itself as a statistical (``accelerated'') ensemble. Using (\ref{Bergman}) we can explicitly compute the partition function as
\be
{\cal Z}({\cal C})=\sum_{j\in\mathbb N/2}\sum^{\infty}_{m=0}\sum^{j}_{q_{1},q_{2}=-j}
 |{\varphi_{q_1,q_2}^{j,m}}({\cal C})|^2=\frac{1}{\det(I-{\cal C}^{\dag}{\cal C})^{\lambda}}=
\frac{1}{(1-\tr({\cal C}^{\dag}{\cal C})+\det({\cal C}^{\dag}{\cal C}))^{\lambda}}.\label{partitionf}\ee
Using this result, the fact that ${\varphi_{q_1,q_2}^{j,m}}({\cal C})$ are homogeneous polynomials of degree $2j+2m$ in ${\cal C}$ (remember Eq. (\ref{energyspectrum}), with
the Hamiltonian operator given by (\ref{Hamiltonian})) and that $\tr({\cal C}^{\dag}{\cal C})$ and $\det({\cal C}^{\dag}{\cal C})$ are homogeneous polynomials of
degree one and two in ${\cal C}$, respectively, the (dimensionless) mean energy in the accelerated ground state (\ref{acvac})
can be calculated as:
 \bea {\cal E}({\cal C})&=&\frac{\sum_{j\in\mathbb N/2}\sum^{\infty}_{m=0}\sum^{j}_{q_{1},q_{2}=-j} E^\lambda_n
|{\varphi_{q_1,q_2}^{j,m}}({\cal C})|^2}{\sum_{j\in\mathbb N/2}\sum^{\infty}_{m=0}\sum^{j}_{q_{1},q_{2}=-j}
 |{\varphi_{q_1,q_2}^{j,m}}({\cal C})|^2}=\lambda\frac{1-\det({\cal C}^\dag {\cal C})}{\det(I-{\cal C}^{\dag}{\cal C})}\nn\\ &=&
{\lambda}+{\frac{-\tr({\cal C}^\dag {\cal C})}{\det(I-{\cal C}^{\dag}{\cal C})}}={\cal E}_0+{{\cal E}_B({\cal C})},\label{meane}
\eea
where we have detached the zero-point (``dark'' energy) contribution ${\cal E}_0=\lambda$ from the rest (``bright'' energy)  ${\cal E}_B({\cal C})$ for convenience.

For the particular case of an acceleration $\alpha$ along the ``$z$'' axis,
${\cal C}=\alpha\sigma^3$, the expressions  (\ref{partitionf}) and (\ref{meane}) acquire the simpler form:
\be
{\cal Z}(\alpha)=(1-\alpha^2)^{-2\lambda},\;\;  {\cal E}(\alpha)=\lambda+2\lambda \frac{\alpha^2}{1-\alpha^2}.\label{partmean}
\ee
Note that the mean energy $ {\cal E}(\alpha)$ is of Planckian type for the identification:
\be \alpha^2(T)\equiv e^{-\frac{\varepsilon}{k_BT}},\label{acceltemp1}\ee
where we have introduced $\varepsilon$ (the quantum of energy
of our four-dimensional harmonic oscillator). At this stage, the identification (\ref{acceltemp1}) is an \textit{ad hoc} assignment but, eventually,
we shall justify it from first thermodynamical principles (see next subsection).

Note also that, for the identification (\ref{acceltemp1}), the partition function ${\cal Z}(\alpha)$ matches that of an Einstein solid with
$2\lambda$ degrees of freedom and Einstein temperature $T_E=\varepsilon/k_B$ (see e.g. \cite{thermo}). We remind the reader that an Einstein
solid consists of $N$ independent (non-coupled) three-dimensional
harmonic oscillators in a lattice (i.e., $\phi=3N$ degrees of freedom). Let us pursue this curious analogy a bit further. The total number of ways to
distribute $n$ quanta of energy among $\phi$
one-dimensional harmonic oscillators is given in general by the binomial
coefficient $W_\phi(n)=\tbinom{n+\phi-1}{\phi-1}$. For example, for $\phi=4$ we recover the degeneracy $W_4(n)=(n+1)(n+2)(n+3)/6$ of
each energy level ${E}_{n}^\lambda$ of our four-dimensional ``conformal oscillator'' given after (\ref{energyspectrum}).
Let us see how  $W_\phi(n)$, for $\phi=2\lambda$, arises from
the distribution function $|\varphi_{q_1,q_2}^{j,m}({\cal C})|^2$. Indeed, for ${\cal C}=\alpha\sigma^3$, $|\varphi_{q_1,q_2}^{j,m}({\cal C})|^2$ can be cast as:
\bea |\varphi_{q_1,q_2}^{j,m}(\alpha)|^2 &=&
\frac{2j+1}{\lambda-1}\binom{m+\lambda-2}{\lambda-2}\binom{m+2j+\lambda-1}{\lambda-2}
(\alpha^2)^{2m}|\cD^{j}_{q_{2},q_{1}}(\alpha\sigma^3)|^2 \nn\\
&=&
\frac{2j+1}{\lambda-1}\binom{m+\lambda-2}{\lambda-2}\binom{m+2j+\lambda-1}{\lambda-2}\alpha^{4j+4m}\delta_{q_1,q_2}\label{probdist1}
 \eea
Fixing $n=2j+2m$,
the  (unnormalized) probability of finding $\tilde\varphi_{0,0}^{0,0}$ in the energy level ${E}_n^\lambda$ is:
\bea
f_n^\lambda(\alpha)&\equiv&\sum^{n/2}_{j=[0,1/2]}\sum^{j}_{q=-j}|\varphi_{q,q}^{j,\frac{n}{2}-j}(\alpha)|^2=
\sum^{n/2}_{j=[0,1/2]}\frac{(2j+1)^2}{\lambda-1}\tbinom{\frac{n}{2}-j+\lambda-2}{\lambda-2}\tbinom{\frac{n}{2}+j+\lambda-1}{\lambda-2}\alpha^{2n}\nn\\
&=&\binom{n+2\lambda-1}{2\lambda-1}\alpha^{2n}=W_{2\lambda}(n)\alpha^{2n},
\label{probdist}\eea
where $[0,1/2]$ is $0$ for $n$  even  and $1/2$ for $n$ odd (in this summation, the $j$ steps are of unity). Here $W_{2\lambda}(n)$ plays the
role of an ``effective'' degeneracy and $\alpha^2$ a Boltzmann-like factor. In fact, the partition function in (\ref{partmean}) can be
obtained again as
\be
{\cal Z}(\alpha)=\sum_{n=0}^\infty f_n^\lambda(\alpha)=\sum_{n=0}^\infty W_{2\lambda}(n)\alpha^{2n}=
\left(\sum_{n=0}^\infty \alpha^{2n}\right)^{2\lambda}=(1-\alpha^2)^{-2\lambda}.\label{part2}
\ee
where we have identified the Maclaurin series expansion of $(1-\alpha^2)^{-2\lambda}$ and the geometric series sum
$z(\alpha)\equiv\sum_{n=0}^\infty \alpha^{2n}=1/(1-\alpha^2)$ with ratio $\alpha^2$. The fact that
${\cal Z}(\alpha)=(z(\alpha))^{2\lambda}$ (the product of $2\lambda$ partition functions $z(\alpha)$) reinforces the analogy
between our accelerated ground state and the Einstein solid with
$2\lambda$ degrees of freedom (see later on next Section for the computation of the entropy).

\begin{figure}[htb]
\begin{center}
\includegraphics[height=5cm,width=12cm]{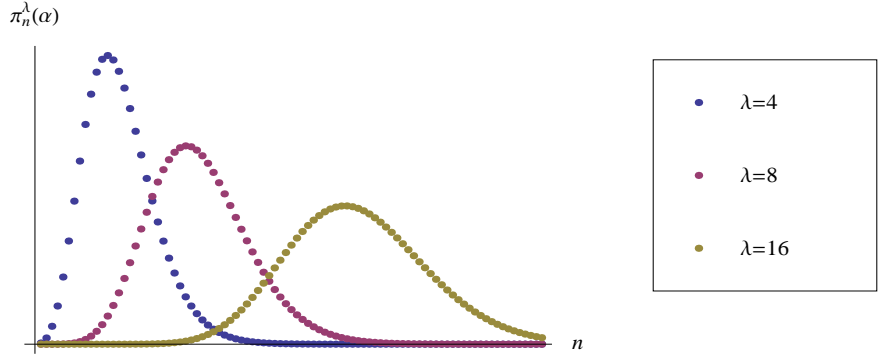}
\end{center}
\caption{Probability $\pi_n^\lambda(\alpha)$ for fixed $\alpha=0.8$ and different values of $\lambda$}\label{n01}
\end{figure}

\begin{figure}[htb]
\begin{center}
\includegraphics[height=5cm,width=12cm]{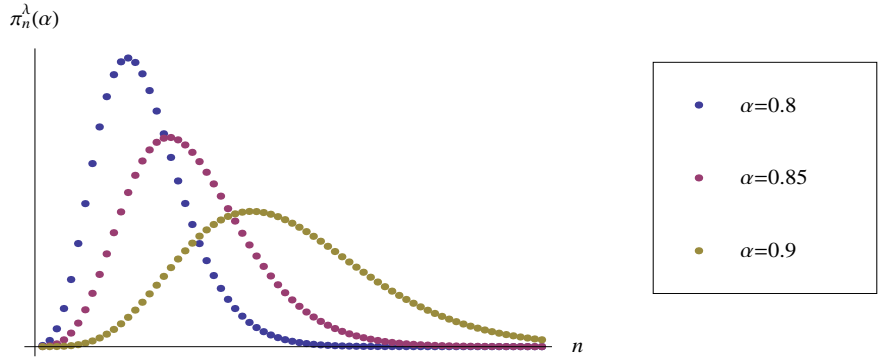}
\end{center}
\caption{Probability $\pi_n^\lambda(\alpha)$ for fixed $\lambda=4$ and different values of $\alpha$}\label{n02}
\end{figure}

Note that the distribution function  $\pi_n^\lambda(\alpha)\equiv f_n^\lambda(\alpha)/{\cal Z}(\alpha)$ has a maximum for a given $n=n_0(\alpha,\lambda)$,
with $n_0(\alpha,\lambda)$ increasing in $\lambda$ (see Figure \ref{n01}) and in $\alpha$ (see Figure \ref{n02}).

\begin{figure}[htb]
\begin{center}
\includegraphics[height=5cm,width=12cm]{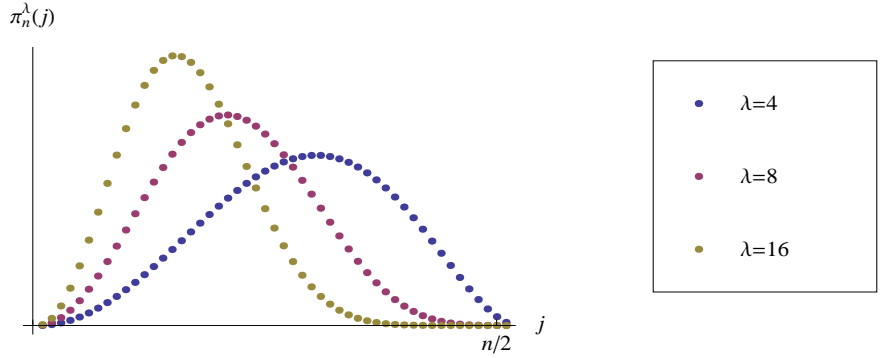}
\end{center}
\caption{Probability $\pi_n^\lambda(j)$ for different values of $\lambda$}\label{j0}
\end{figure}

Furthermore, inside each energy level ${E}_n^\lambda$, the allowed angular momenta $j=[0,1/2],\dots,n/2$ appear with different (unnormalized) probabilities:
\be
f_{n,j}^\lambda(\alpha)\equiv\frac{(2j+1)^2}{\lambda-1}\binom{\frac{n}{2}-j+\lambda-2}{\lambda-2}\binom{\frac{n}{2}+j+\lambda-1}{\lambda-2}\alpha^{2n}.
\ee
Actually, the distribution function $\pi_n^\lambda(j)\equiv f_{n,j}^\lambda(\alpha)/f_n^\lambda(\alpha)$, which is independent of $\alpha$, has a maximum for a given $j=j_0(n,\lambda)$,
with $j_0(n,\lambda)$ an increasing sequence of $n$ and decreasing on $\lambda$ (see Figure \ref{j0}).

\subsection{Entropy, temperature and ¿maximal acceleration?}

Note that, deriving the partition function ${\cal Z}(\alpha)$
and mean energy $ {\cal E}(\alpha)$ from the distribution function (\ref{probdist1},\ref{probdist})
does not involve any thermal (but just pure quantum mechanical) input. In the same way, we can also compute
the entropy as a logarithmic measure of the density of states. In fact, denoting by
$p_n(\alpha)=\alpha^{2n}/{\cal Z}(\alpha)$ the probability of finding our ``Einstein solid'' in the energy level $n$ with degeneracy
$W_\lambda(n)$, the entropy can be calculated as
\bea {\cal S}(\alpha)&=&-\sum_{n=0}^\infty W_{2\lambda}(n) p_n(\alpha)\ln
p_n(\alpha)\nn\\ &=&-\sum_{n=0}^\infty\tbinom{2\lambda+n-1}{n}
(1-\alpha^2)^{2\lambda}\alpha^{2n}\ln((1-\alpha^2)^{2\lambda}
\alpha^{2n})
\nn\\ &=&-(1-\alpha^2)^{2\lambda}\left(\sum_{n=0}^\infty
\tbinom{2\lambda+n-1}{n}\alpha^{2n}\ln((1-\alpha^2)^{2\lambda})
+\sum_{n=0}^\infty\tbinom{2\lambda+n-1}{n}\alpha^{2n}\ln(
\alpha^{2n})\right)\nn\\ &=&-(1-\alpha^2)^{2\lambda}\left(2\lambda\ln(1-\alpha^2)\sum_{n=0}^\infty
\tbinom{2\lambda+n-1}{n}\alpha^{2n}
 +2\ln(\alpha)\sum_{n=1}^\infty\tbinom{2\lambda+n-1}{n}n\alpha^{2n}\right),\nn\\ 
&=& -2\lambda\left(\frac{\alpha^2\ln(\alpha^2)}{1-\alpha^2}+\ln(1-\alpha^2)\right), \label{proofentropy}\eea
where we have identified the partition function ${\cal Z}(\alpha)$ and its derivative $\alpha^2\frac{d}{d(\alpha^2)}{\cal Z}(\alpha)$ in the last 
two summations. Again, there is not any thermal input up to now. If we wanted to assign an ``effective'' temperature ${\cal T}$ to our ``accelerated ensemble'',
we could use the universal thermodynamic expression (derivative of the energy with respect to the entropy):
\be
{\cal T}=\frac{d {\cal E}(\alpha)}{d {\cal S}(\alpha)}= -\frac{1}{\ln(\alpha^2)}, \label{tempac}
\ee
given in unities of the Einstein temperature $T_E$ previously introduced (i.e. ${\cal T}=T/T_E$).\footnote{The semisimple character of the group
$SU(2,2)$ allows us to express all kinematic magnitudes by pure numbers.  From a ``Galilean'' viewpoint, we could say that in  conformal
kinematics there is a characteristic length, a characteristic time and a characteristic
speed which may be used as natural units, and then lengths, times and speeds are
dimensionless (see \cite{group-dim1,group-dim2} for a thorough study on kinematic groups and dimensional analysis)}. The equality (\ref{tempac})
can be inverted to formula (\ref{acceltemp1}), giving the announced  derivation of the assignment (\ref{acceltemp1}) from first thermodynamic principles.
One could still check \textit{{consistency}} (if desired)  with other classical formulas relating mean energy and entropy to the partition function, namely:
\be {\cal E}(\alpha)=-\frac{d\ln{\cal Z}(\alpha)}{d\beta}, \;\;{\cal S}(\alpha)=\frac{d}{d{\cal T}}({\cal T}\ln{\cal Z}(\alpha)),\; \beta\equiv 1/{\cal T}.
\ee
A hurried analysis of the relation $\alpha^2=e^{-1/{\cal T}}$ would lead to think of the existence of a ``maximal acceleration'' $\alpha^2=1$ (in dimensionless 
unities). Actually, in the process towards the calculation of thermodynamical quantities, we have made use of a rescaling of the original acceleration 
$C=c_\mu\sigma^\mu$, in the expression (\ref{acvac1}), to
${\cal C}=D^{-1}C=\alpha_\mu\sigma^\mu$. We can find the relation between $c_\mu$ and $\alpha_\mu$ as follows. 
Taking int account that $\varphi^{00}_{00}$ is normalized and the representation (\ref{repre}) is unitary (see Appendix C and Proposition 5.2 of \cite{MacMahon}), 
we know that the accelerated ground state (\ref{acvac1}) is also normalized. This means that the normalizing global factor $\det(D)^{-\lambda}$ in (\ref{acvac1}) 
is related to the partition function ${\cal Z}({\cal C})$ in (\ref{partitionf}) by:
\be
\det(DD^\dag)^{-\lambda}=1/{\cal Z}({\cal C})=\det(I-{\cal C}^{\dag}{\cal C})^{\lambda}\Rightarrow \det(DD^\dag)=\frac{1}{\det(I-{\cal C}^{\dag}{\cal C})}.\ee
Therefore, for $C=c\sigma^3$ and ${\cal C}=\alpha\sigma^3$, the relation ${\cal C}^\dag{\cal C}=C^\dag(DD^\dag)^{-1}C$ reads: 
\be
\alpha^2=\frac{c^2}{1+c^2}\Rightarrow c^2=\frac{\alpha^2}{1-\alpha^2}.\ee
With this identification, the mean energy ${\cal E}=\lambda+2\lambda \frac{\alpha^2}{1-\alpha^2}=\lambda+2\lambda c^2$ turns out to be a 
\textit{quadratic} function of the acceleration $c$. The dependence of $c$ with the effective temperature ${\cal T}$ is then:
\be
c=\sqrt{\frac{e^{-1/{\cal T}}}{1-e^{-1/{\cal T}}}}= \sqrt{{\cal T}}+O(\frac{1}{\sqrt{{\cal T}}}) \,\,{\rm for}\,\, {\cal T}>>1.\ee
This behavior departs from the Unruh's formula (\ref{tempunruh}) even in the limit of  high temperatures.

We have seen that the fact that $\alpha$ is bounded is just due to a rescaling of $c$, so that there is not a maximal acceleration in our model as such. 
Nevertheless,  we would like to comment on other arguments in the literature supporting
the existence of a bound $a_{\rm max}$ for proper accelerations. One  was given time
ago in Ref. \cite{conforme-maxac} in connection with conformal kinematics; there the authors analyzed the
physical interpretation of the singularities, $1+2c x+c^2 x^2=0$,
of the SCT  (\ref{confact-1}). When applying the
transformation to an extended object of size $\ell$, an
upper-limit to the proper acceleration, $a_{\rm max}\simeq
v^2/\ell$, is shown to be necessary in order to the tenets of
special relativity not to be violated (see \cite{conforme-maxac} for
more details). Before, Caianiello \cite{Caianiello} derived the existence and physical
consequences of a maximal acceleration connected with the Born's Reciprocity Principle (BPR) \cite{Born1,Born2}. 
Indeed, one can deduce the existence of a maximal
acceleration from the positivity of the Born's line element
\be d\tilde\tau^2=dx_\mu dx^\mu+\frac{\ell^4_{\rm
min}}{\hbar^2}dp_\mu dp^\mu=d\tau\sqrt{1-\frac{|a^2|}{a^2_{\rm
max}}}\label{Bornline}\ee
where $d\tau^2\equiv dx_\mu dx^\mu$ and $dp_\mu/d\tau\equiv m
d^2x_\mu/d\tau^2=ma_\mu$, as usual.
An adaptation of the BRP to the conformal relativity has been put forward by some of us in \cite{spinning-part}, where a conformal analogue
of the line element (\ref{Bornline}) in the phase space $\mathbb D_4$ has been considered. However, the existence of a maximal acceleration inside
the conformal group does not seem to be apparent neither from this conformal adaptation of the BRP.

In the last years, many papers have been published  (see e.g. \cite{Castro2} and references therein),
each one introducing the maximal acceleration starting from
different motivations and from different theoretical schemes.
Among the large list of physical applications of Caianiello's
model we would like to point out the one in cosmology which avoids
an initial singularity while preserving inflation. Also, a
maximal-acceleration relativity principle leads to a variable fine
structure ``constant''  \cite{Castro2}, according to which it
could have been extremely small (zero) in the early
Universe and then all matter in the Universe could have emerged
via the Unruh effect. Moreover, in a non-commutative geometry setting \cite{Madore}, the
non-vanishing commutators (\ref{noncomspace}) can be seen as a
sign of the granularity (non-commutativity) of space-time in
conformal-invariant theories, along with the existence of a
minimal length $\ell_{\rm min}$ or, equivalently, a maximal acceleration $a_{\rm max}=v^2/\ell_{\rm min}$.

\section{Second-Quantized Theory, Conformal Zero Modes and Poincaré $\theta$-vacua\label{thetasec}}

We have discussed the effect of relativistic accelerations in first quantization. However, the proper setting to analyze radiation
effects is in the second-quantized theory. Let us denote (for space-saving
notation) by $n=\{j,m,q_1,q_2\}$ the multi-index of the one-particle basis wavefunctions $\varphi_n$ in
(\ref{polibase}) and by $\hat a_n$ (resp. $\hat a^\dag_n$) operators annihilating (resp. creating) a particle
in the state $|n\rangle$. As for the case of a single bosonic mode in (\ref{Fockbasis}), an orthonormal basis for the Hilbert space of the second-quantized
theory is constructed by taking the orbit through the \textit{conformal vacuum} 
$|{0}\rangle$ of the creation operators $\hat a^\dag_n$:
\be |q(n_1), \dots,
q(n_p)\rangle\equiv\frac{(\hat a^\dag_{n_1})^{q(n_1)}\dots
(\hat a^\dag_{n_p})^{q(n_p)} }{(q(n_1)!\dots
q(n_p)!)^{1/2}}|{0}\rangle,\label{orbit1} \ee
where $q(n)\in \mathbb N$ denotes the occupation number of the state $n$ with
energy  $2j+2m$.

The fact that the ground state of the first
quantization, $\varphi_0$, is invariant under Poincaré transformations (remember the discussion after (\ref{acvac})),  
implies that  the annihilation operator $\hat a_0$ of zero-(``dark'')-energy modes commutes with all Poincaré generators. It also
commutes with all annihilation operators and creation operators of particles with positive (``bright'') energy,
\be [\hat a_0,\hat a_n^\dag]=0,\; n\not=0.\ee
Therefore, by Schur's Lemma, $\hat a_0$
must behave as a multiple of the identity when conformal symmetry is broken/restricted to Poincaré symmetry. This means 
that we can choose Poincaré-invariant vacua $|\theta\rangle$ as being eigenstates of $\hat a_0$, namely:
\be \hat a_0|\theta\rangle=\theta|\theta\rangle\Rightarrow {
|\theta\rangle= e^{\theta \hat a^\dag_0-\bar\theta \hat a_0}|0\rangle},\ee
which implies that Poincaré ``$\theta$-vacua''  $|\theta\rangle$  are (canonical) \emph{{coherent states of
conformal zero modes}} (remember the general definition in Eq. (\ref{exp-ansion})). Unlike the conformal vacuum $|0\rangle$, which is invariant 
under the whole conformal group, Poincaré $\theta$-vacua  $|\theta\rangle$ are not stable under special conformal transformations (accelerations). 
In fact, the second-quantized version of (\ref{acvac1}), for an acceleration
${\cal C}=\alpha\sigma^3$ along the third axis, is given by the transformation of annihilation (resp. creation) operators:
\be\tilde{\hat{a}}_0=\sum_{n=0}^\infty \varphi_n(\alpha)\hat a_n.\ee
We shall assume that $\sum_n|\varphi_n(\alpha)|^2=1$ (normalized probabilities) so that this transformation preserves the 
original commutation relations $[\tilde{\hat{a}}_0,\tilde{\hat{a}}_0^\dag]=1$. Therefore, accelerated Poincaré $\theta$-vacua are:
\be|\tilde\theta\rangle=e^{\theta \tilde{\hat{a}}^\dag_0-\bar\theta\tilde{\hat{a}}_0}|0\rangle,\label{actvac}\ee
which can also be written as
\be|\tilde\theta\rangle=e^{\theta \sum_{n= 1}^\infty \overline{\varphi_n(\alpha)}\,\tilde{\hat{a}}^\dag_n}|\theta\rangle.\label{actvac2}\ee
We can think of conformal zero modes as ``virtual particles'' without ``bright'' energy and undetectable by inertial observers. However, 
from an accelerated frame, they become ``visible'' to a Poincaré observer. The average number of particles with energy $E_n$ in the accelerated
vacuum (\ref{actvac}) is then given by
\be N_n(\alpha)=\langle\tilde\theta|\hat a^\dag_n \hat
a_n|\tilde\theta\rangle=|\theta|^2|\varphi_n(\alpha)|^2,\ee
where $|\theta|^2$ is the total average number of particles in
$|\theta\rangle$, and $|\varphi_n(\alpha)|^2$ is the occupation number of the energy level $E_n$ of the accelerated vacuum
$|\tilde\theta\rangle$. The situation resembles that in many condensed-matter systems 
(like Bose-Einstein condensates, superconductors, etc), where one also finds  non-empty, coherent 
ground states.

In the same way, the probability
$P_n(q,\alpha)$ of observing $q$ particles with energy $E_n$ in
$|\tilde\theta\ra$ can be calculated as:
\be P_n(q,\alpha)=|\la q(n)
|\tilde\theta\ra|^2=\frac{e^{-|\theta|^2}}{q!}|\theta|^{2q} \,
|\varphi_n(\alpha)|^{2q}=\frac{e^{-|\theta|^2}}{q!}N_n^{q}(\alpha).\label{probpart}
\ee
Therefore, the relative probability of observing a state with
total energy $E$ in the excited vacuum $|\tilde\theta\rangle$ is:
\be P(E)=\sum_{\begin{array}{c} q_0,\dots,q_k :\\ \sum^k_{n=0}E_n
q_n=E\end{array}} \prod^k_{n=0}P_n({q}_n,\alpha)\,.\label{probest}\ee
For the case studied in this paper, this distribution function
can be factorized as $P(E)=\Omega(E)e^{-E/{\cal T}}$, where
$\Omega(E)$ is a relative weight proportional to the number of
states with energy $E$ and the factor $e^{-E/{\cal T}}$ fits this
weight properly to a temperature ${\cal T}$.

One can also compute the  total mean energy
\be {E}(\alpha)=\langle\tilde\theta|\sum_{n=1}^\infty E_n\hat a^\dag_n\hat a_n|\tilde\theta\rangle=|\theta|^2 \sum_{n=1} |\varphi_n(\alpha)|^2
E_n=|\theta|^2{\cal E}(\alpha),\label{meanenergy}\ee
which, as expected, is the product of ${\cal E}(\alpha)$ by the average number of particles $|\theta|^2$ in $|\theta\rangle$.  The free parameter  
$|\theta|^2$ is also linked to a vacuum (``dark'') energy $E_0=|\theta|^2{\cal E}_0=|\theta|^2\lambda$ whose value should be determined by experiments, just like,
for example, the cosmological constant.  Like other non-zero vacuum expectation values,  zero-point energy leads to observable consequences as, for 
instance, the Casimir effect, and influences the behavior of the Universe at cosmological scales, where the vacuum (dark) energy is 
expected to contribute to the cosmological constant, which affects the expansion of the universe (see e.g. \cite{vacmith} for a nice review). Actually, dark energy 
is the most popular way to explain recent observations that the universe appears to be expanding at an accelerating rate.

\section{Comments and Outlook}\label{conclu}

As already commented in the Introduction,  conformal field theories also seem to
provide a universal  description of low-energy black hole thermodynamics, which
is only fixed by symmetry arguments  (see \cite{Carlipcqg,Agullo} and references therein).
Actually, Unruh's temperature
(\ref{tempunruh}) coincides with Hawking's temperature
\begin{equation}
T= \frac{\hbar v^3}{8\pi M k_B G}=\frac{2\pi GM\hbar}{\Sigma v k_B}
\label{tempHawking}
\end{equation}
($\Sigma=4\pi r_g^2=
8\pi G^2M^2/v^4$ stands for the surface of the event horizon)
when the acceleration is that of a free falling observer on the surface
$\Sigma$, i.e.  $a=v^4/(4GM)=GM/r_g^2$. Here,
the Virasoro algebra proves to be a physically important
subalgebra of the gauge algebra of surface deformations that leave
the horizon fixed for an arbitrary black hole. Thus, the fields on the surface
must transform according to irreducible representations of the Virasoro
algebra, which is the general symmetry principle that
governs the density of microscopic states. Bekenstein-Hawking expression
for the entropy can be then calculated from the Cardy formula
\cite{Cardy,Cardy2} (see also \cite{Carlip} for logarithmic corrections).
Therefore, in the Hawking effect,  the calculation of thermodynamical quantities, linked to the statistical
mechanical problem of counting microscopic states,  is reduced
to the study of the representation theory of the conformal group.

Although our approach to the quantum analysis of accelerated frames shares with the previous description of black hole thermodynamics the existence 
of an underlying conformal invariance, we should not confuse both schemes. Conformal invariance in Hawking effect manifests itself as an infinite-dimensional  
gauge algebra of (two-dimensional) surface deformations. However, the infinite-dimensional character of conformal symmetry seems to be an exclusive
patrimony of two-dimensional physics, and conformal invariance in (3+1)-dimensions is finite-(15)-dimensional, thus accounting for 
transitions to uniformly accelerated frames only.  To account for 
higher-order effects of gravity on quantum field theory  from a group-theoretical point of view, 
one should consider more general diffeomorphism (Lie) algebras.   Higher-dimensional analogies of
the infinite two-dimensional conformal symmetry have been proposed by us in  \cite{infdimal,promoting,CQG,gp,jarama}. We think that these inifinite ${\cal W}$-like symmetries
can play some fundamental role in quantum gravity models, as a gauge guiding principle.

To conclude, we would also like to mention that, the same spontaneous $SU(2,2)$-symmetry breaking mechanism explained in this paper applies to general
$SU(N,M)$-invariant quantum  theories, where an interesting
connection between ``curvature and statistics'' has emerged
\cite{curvstat,vacrad}. We hope that many more interesting
physical phenomena  remain to be unraveled inside
conformal-invariant quantum (field) theory.

\section*{Acknowledgements}
Work partially supported by the Fundación Séneca (08814/PI/08),
Spanish MICINN (FIS2008-06078-C03-01) and Junta de Andaluc\'\i a (FQM219,
FQM1951).  M. Calixto thanks the ``Universidad Politécnica de
Cartagena'' and C.A.R.M.  for the award  ``Intensificación de la
Actividad Investigadora 2009-2010''. ``Este trabajo es resultado de la
ayuda concedida por la Fundación Séneca, en el marco del PCTRM
2007-2010, con financiación del INFO y del FEDER de hasta un 80\%''.

\appendix

\section{Vacuum radiation as a consequence of space-time mutilation\label{mutilation}}

The  existence of event horizons in passing to accelerated frames
of reference leads to unitarily inequivalent representations of
the quantum field canonical commutation relations and to a
(in-)definition of particles depending on the state of motion of the
observer.


To use an explicit example, let us consider a real scalar massless field
$\phi(x)$, satisfying the Klein-Gordon equation \be
\eta^{\mu\nu}\partial_{\mu}\partial_\nu \phi(x)=0 \, .\label{kg} \ee Let
us denote by $a_k,a^*_k$  the Fourier coefficients of the decomposition of
$\phi$ into positive and negative frequency modes:
\be \phi(x)=\int dk( a_kf_k(x)+ a^*_k f^*_k(x)). \ee
The Fourier coefficients $a_k,a^*_k$ are promoted to annihilation and
creation operators $\hat a_k, \hat a^*_k$ of particles in the quantum
field theory. The Minkowski vacuum $|0\rangle_M$ is defined as the state
nullified by all annihilation operators
\be \hat a_k|0\rangle_M=0,\;\forall k.\label{vacM}\ee


Let us consider now the Rindler coordinate transformation (see e.g.
\cite{Birrell}):
 \be
t=a^{-1}e^{az'}\sinh(at'),\;\;
z=a^{-1}e^{az'}\cosh(at').\label{Rindlert}\ee
The worldline $z'=0$ has constant acceleration $a$ (in natural unities).
This transformation entails a mutilation of Minkowski spacetime into
patches or charts with event horizons.

The new coordinate system provides a new decomposition of $\phi$ into
Rindler positive and negative frequency modes:
\be \phi(x')=\int dq(a'_qf'_q(x')+ a'^*_q f'^*_q(x')). \ee
The Rindler vacuum $|0\rangle_R$ is defined as the state nullified by all
Rindler annihilation operators: \be \hat a'_q|0\rangle_R=0\,\forall
q.\label{vacR}\ee

One can see that the Minkowski  vacuum  $|0\rangle_M$ and the Rindler vacuum
$|0\rangle_R$ are not identical. Actually, the Minkowski vacuum
$|0\rangle_M$  has a nontrivial content of Rindler particles. In fact, 
the Fourier components $a'_q, {a'}^*_q$ of the field $\phi$ in the new
(accelerated) reference frame are expressed in terms of both $a_k,a^*_k$
through a Bogolyubov transformation:
 \bea
&a'_q=\int{dk\left(\alpha_{qk} a_k+\beta_{qk} a^*_k\right)}\,, &\nn\\
& \alpha_{qk}=\langle f'_q|f_k\rangle,\;\;\beta_{qk}=\langle
f'_q|f^*_k\rangle.& \label{bog}\eea
The vacuum states $|0\rangle_M$ and $|0\rangle_R$, defined by the
conditions (\ref{vacM}) and (\ref{vacR}), are not identical if the
coefficients $\beta_{qk}$ in (\ref{bog}) are not zero. In this case the
Minkowski vacuum has a non-zero average number of Rindler particles given
by:
\be N_R=\langle 0|\hat N_R|0\rangle_M=\langle 0|\int dq \hat a'^\dag_q\hat
a'_q |0\rangle_M=\int{dkdq|\beta_{qk}|^2}.\ee
That is,  both quantizations are inequivalent.

\section{Conformal particles from constrained nonlinear $\sigma$-models on SU(2,2)\label{conf-sigma}}

Let us briefly report on a 
$G_0\equiv SL(2,\mathbb C)\times \mathbb R^+$ (Lorentz times dilations) gauge-invariant
Lagrangian approach (of sigma-model type) to the formulation of conformal $G=SU(2,2)$ invariant quantum particles, with the cotangent of $G$ as a preliminary phase space. 
This formulation has been
thoroughly developed in Reference \cite{spinning-part}, where we used a generalized Dirac
method for the quantization of constrained systems, which resembles in some aspects
the standard approach to quantizing coadjoint orbits of a group $G$ (see e.g. classical References as \cite{Kostant}, \cite{Kirillov} and \cite{Bal}). 

Denote by
$g(t)\in G$ trajectories on $G$, $\vartheta^L\equiv-ig^{-1} dg=-ig^{-1} \dot g dt$ the left-invariant Maurer-Cartan one-form and $\Gamma$ the hermitian 
form in Eq. (\ref{su22def}). Then the singular action
\be A(g,\dot g)=\int\tr(\Gamma\vartheta^L)\label{sigmaaction}\ee
is naturally left-$G$-invariant under rigid transformations $g(t)\to g' g(t), \forall g'\in G$, the infinitesimal generators of this symmetry  
(right-invariant  vector fields) being the basic operators/observables of the first-quantized theory. In addition, the action (\ref{sigmaaction}) is also 
right-$G_0$-invariant  under local-gauge transformations $g(t)\to g(t)g_0(t), \;\forall g_0(t)\in G_0$ (see \cite{spinning-part} for a proof).
At the quantum level, the $G_0$ gauge right-invariance of the
proposed Lagrangian manifests itself by leaving  complex wave functions on $G$, $\phi:G\to\mathbb C$, right-invariant under gauge-group $G_0$ transformations,
i.e. $\phi(gg_0)=\phi(g), \forall g_0\in G_0, g\in G$. Actually, the last strict invariant condition,  $\phi(gg_0)=\phi(g)$, can be relaxed to
``invariance up to a phase'', $\phi(gg_0)=u_l(g_0)\phi(g), u_s(g_0)\in U(1)$,
thus allowing internal degrees of freedom, like the spin $l=s$, to enter the theory ($l$ denotes any label characterizing the representation). 
The ``Gauss-law-like'' constraints related to 
this $G_0$ gauge right-invariance are written in terms of the corresponding infinitesimal generators  (left-invariant `$L$' vector fields) $M_{\mu\nu}^L$ and 
$D^L$ of Lorentz and dilation transformations as:  $M_{\mu\nu}^L\phi=0$ (for spin-less $s=0$ particles) and $D^L\phi=\lambda\phi$ (for conformal, scale or 
mass dimension $\lambda$), respectively. These constraints restrict the support of wave functions $\phi$ from $G$ to the eight-dimensional domain 
$\mathbb D_4=G/G_0$ of the complex Minkowski  (phase) space $\mathbb C^4$.  An extra condition
$K_\mu^L\phi=0$ is just intended to select the ``position'' (versus ``momenta'') representation and it is necessary for irreducibility of the quantum representation. 
The last condition further restricts the support of $\phi$ to $\mathbb R^4$. This means that Cauchy hypersurfaces
have dimension four. In other words, the time-translations generator $P_0$ is now a dynamical
operator, on an equal footing with spatial-translations generators $P_j$, thus suffering Heisenberg indeterminacy relations
too. This fact can also be inferred from the generalized Klein-Gordon equation $C_2^L\phi=m_{00}^2\phi$, with $C^L_2$ the quadratic conformal Casimir 
(\ref{Casimir}) (which has the same expression in terms of left-invariant generators as in terms of right-invariant ones) and $m_{00}^2=\lambda(\lambda+4)$ 
(indeed, use the constraints $M_{\mu\nu}^L\phi=0$, $K_\mu^L\phi=0$ and $D^L\phi=\lambda\phi$ and the last commutation relation in (\ref{conformalgebra})). 
In fact, let us consider the alternative (vector and pseudo-vector) combinations
\[\tilde P_\mu\equiv\um(P_\mu+K_\mu),\;\;\tilde
K_\mu\equiv\um(P_\mu-K_\mu),\]
with new commutation relations:
\be\left[\tilde P_\mu,\tilde
K_\nu\right]=\eta_{\mu\nu}D,\;\left[\tilde P_\mu,\tilde
P_\nu\right]=M_{\mu\nu},\; \left[\tilde K_\mu,\tilde
K_\nu\right]=-M_{\mu\nu},\label{noncomspace}\ee
in terms of which the Casimir (\ref{Casimir}) reads
\be C_2=D^2-\um M_{\mu\nu}M^{\mu\nu}+\tilde P_\mu \tilde P^\mu-\tilde K_\mu \tilde K^\mu.\label{Casimir2}\ee
A new compatible set of constraints are: $M_{\mu\nu}^L\phi=0$ and 
$\tilde K^L_\mu\phi=0$,  resulting in wave functions $\phi(g)$, having support on the extended 
Minkowski space $G/{\cal P}\simeq \mathbb R^4\times \mathbb R^+$. The commutator $[D^L,\tilde K^L_\mu]=-\tilde P^L_\mu$ now 
precludes the imposition of $D^L\phi=\lambda\phi$ (otherwise we would be forced to impose $\tilde P^L_\mu\phi=0$ too and the representation would 
be trivial). Instead, the Casimir constraint 
\be C_2^L\phi=m_{00}^2\phi \Rightarrow ((D^L)^2+(\tilde P^L)^2)\phi=m_{00}^2\phi\label{genKG}\ee
can be compatibly imposed,  togheter with  $M_{\mu\nu}^L\phi=0$ and 
$\tilde K^L_\mu\phi=0$, thus leading to the announced generalized Klein-Gordon equation with $D$ playing the role of the new (proper) time 
evolution generator.

\end{document}